\documentclass[a4paper,10pt,twoside]{cpc-hepnp}

\usepackage{multicol}
\usepackage{graphicx}
\usepackage{booktabs}
\usepackage{amssymb,bm,mathrsfs,bbm,amscd}
\usepackage[tbtags]{amsmath}
\usepackage{lastpage}
\usepackage{CJK}
\usepackage{color}

\begin{document}
\begin{CJK*}{GB}{gbsn}

\fancyhead[c]{\small Chinese Physics C~~~Vol. xx, No. x (201x) xxxxxx}

\footnotetext[0]{Received xxxx}

\title{ Behavior and finite-size effects of the sixth order cumulant in the three-dimensional Ising universality class\thanks{Supported by Fund Project of Sichuan Provincial Department of Education (16ZB0339), Fund Project of Chengdu Technological University for Doctor (2016RC004), the Major State Basic Research Development Program of China (2014CB845402) and National Natural Science Foundation of China (11405088, 11221504)}}

\author{%
      Xue Pan $^{1;1)}$\email{panxue1624@163.com}%
      \quad Li-Zhu Chen $^{2}$%
\quad Yuan-Fang Wu $^{3}$%
}
\maketitle

\address{%
$^1$ School of Electronic Engineering, Chengdu Technological University, Chengdu 611730, China\\
$^2$ School of physics and Optoelectronic Engineering, Nanjing University of Information Science and Technology, Nanjing 210044, China\\
$^3$ Key Laboratory of Quark and Lepton Physics (MOE) and
Institute of Particle Physics, Central China Normal University, Wuhan 430079, China\\
}

\begin{abstract}
The high-order cumulants of conserved charges are suggested to be sensitive observables to search for the critical point of Quantum Chromodynamics (QCD). This has been calculated to the sixth order in experiments. Corresponding theoretical studies on the sixth order cumulant are necessary. Based on the universality of the critical behavior, we study the temperature dependence of the sixth order cumulant of the order parameter using the parametric representation of the three-dimensional Ising model, which is expected to be in the same universality class as QCD. The density plot of the sign of the sixth order cumulant is shown on the temperature and external magnetic field plane. We found that at non-zero external magnetic field, when the critical point is approached from the crossover side, the sixth order cumulant has a negative valley. The width of the negative valley narrows with decreasing external field. Qualitatively, the trend is similar to the result of Monte Carlo simulation on a finite-size system. Quantitatively, the temperature of the sign change is different. Through Monte Carlo simulation of the Ising model, we calculated the sixth order cumulant of different sizes of systems. We discuss the finite-size effects on the temperature at which the cumulant changes sign.

\end{abstract}

\begin{keyword}
critical point, the sixth order cumulant, Ising model
\end{keyword}

\begin{pacs}
25.75.Gz, 25.75.Nq
\end{pacs}

\footnotetext[0]{\hspace*{-3mm}\raisebox{0.3ex}{$\scriptstyle\copyright$}2013
Chinese Physical Society and the Institute of High Energy Physics
of the Chinese Academy of Sciences and the Institute
of Modern Physics of the Chinese Academy of Sciences and IOP Publishing Ltd}%

\begin{multicols}{2}

\section{Introduction}

Quantum Chromodynamics (QCD) is generally believed to be the theory of describing the strong interaction in terms of quarks and gluons. It is expected that the QCD system undergoes a first order phase transition from hadronic matter to quark-gluon plasma (QGP) at high baryon density and low temperature~\cite{first-PRD, first-NPB,first-PRL,first-NPB1,first-LQCD,first-PRC}. The first order phase transition ends at a critical point, which is a unique character of the QCD phase diagram on the baryon chemical potential-temperature ($\mu_B-T$) plane. Beyond the critical point, it is expected to be a rapid crossover. This has been proved at vanishing baryon chemical potential by lattice QCD~\cite{fodor-nature}. One of the main goals of heavy-ion collision experiments is to locate the QCD critical point and determine the energy scale of QCD phase transitions.

In the thermodynamic limit, the correlation length ($\xi$) is infinite at the critical point. Susceptibility should diverge as $\xi^2$. But in a finite system, $\xi$ does not diverge and the susceptibility is rounded to a finite peak~\cite{round1, round2}. As a system created in heavy-ion collisions, the magnitude of the correlation length is also limited by the system size but mainly limited by the finite-size effects due to critical slowing down~\cite{length1, length2}. A more sensitive observable than the susceptibility is needed in order to search for the critical point.

In recent years, the high-order cumulants (this is a mathematical term, corresponding to generalized susceptibilities in physics) of conserved charges are suggested to be sensitive observables to search for the QCD critical point in  relativistic heavy-ion collisions, e.g., the net baryon number, net electric charge, and net strangeness~\cite{stephanov-prl91, koch, Stephanov-prl102, Asakawa-prl103, Stephanov-prl107, Karsch-EPJC71}. They are particularly sensitive to the correlation length ($\xi$). For example, the third order cumulant diverges with $\xi^{4.5}$, and the fourth order cumulant diverges with $\xi^7$~\cite{Stephanov-prl102}.

There are already many works on the high-order cumulants of conserved charges from both experiments and theory~\cite{Asakawa-prl103, Stephanov-prl107,Karsch-EPJC71, Mcheng, Fuweijie, Vladi, Vladi1, Vladi2, STAR}. The theoretical results show there exists non-monotonic or sign change behavior in the vicinity of the critical point in the high-order cumulants. In Ref.~\cite{Karsch-PLB695}, the authors found that the behavior of the first four orders of cumulants of net-proton number at the Relative Heavy-Ion Collider (RHIC) calculated by the STAR Collaboration~\cite{STAR} agrees well with the hadron resonance gas (HRG) model and lattice QCD calculations. Based on the three-dimensional $O(4)$ scaling function, the sixth order cumulant of baryon number deviates considerably from the HRG model. It remains negative at the chiral transition temperature~\cite{Karsch-EPJC71}. The preliminary results of the sixth order cumulant in STAR at RHIC have been shown in Ref.~\cite{chenlz-NPA}. It is necessary to study the behavior of the sixth order cumulant from theory.

If the QCD critical point exists, it belongs to the same universality class as the three-dimensional Ising model~\cite{class 1,class 2, class 3,class 4}. They have the same values for the critical exponents. The behavior of the corresponding thermodynamic quantities is similar. That is to say the sixth order cumulant of the order parameter in the three-dimensional Ising model can reflect the properties of the sixth order cumulant of the net-baryon.

In Ref.~\cite{NPA}, through the Monte Carlo simulation of the Ising model, we have studied the temperature dependence of the sixth order cumulant of the order parameter at three fixed external magnetic fields in a finite system. We found that when approaching the critical point from the crossover side, the sixth order cumulant is negative.

In this paper, using the parametric representation, we further study the distribution of the sign of the sixth order cumulant of the order parameter in the temperature-external magnetic field ($t-H$) plane, and its temperature dependence at a fixed external field in the thermodynamic limit. The qualitative behavior is compared to the results from the Monte Carlo simulations.

As the magnitude of $\xi$ is limited in the system created in heavy ion collisions, the finite-size effects cannot be neglected. In a system of finite-size, we  can just get the pseudo-critical point. Within the linear sigma model with constituent quarks and a two-flavor quark-meson model, the authors found that the pseudo-critical point decreases in the $T-\mu$ plane as the system size decreases~\cite{finitesize1, finitesize2, finitesize3, finitesize4}. In this paper, we study the finite-size effects on the temperature at which the sixth order cumulant changes sign on different sizes of lattice by Monte Carlo simulations.

The paper is organized as follows. In Section 2, the Ising model and its parametric representation are introduced. The parametric expression of the sixth order cumulant is derived. In Section 3, the distribution of the sign of the sixth order cumulant is presented. Its temperature dependence at a fixed external magnetic field is compared with the simulation results of the Monte Carlo method from the three-dimensional Ising model. In Section 4, the finite-size effects on the sixth order cumulant are discussed. The results of the sixth order cumulant are compared with the fourth order one. Finally, the conclusions and summary are given in Section 5.

\section{The Ising model and parametric expression of the sixth order cumulant}

The three-dimensional Ising model is defined as follows,
\begin{flalign}\label{Ising model}
&\qquad \mathcal{H} =-J\sum_{\langle i,j\rangle}{s}_{i}{s}_{j}-H\sum_{i}{s}_{i},&
\end{flalign}
where $\mathcal{H}$ is the Hamiltonian, and $s_{i}$ is the spin at site $i$ on a cubic lattice which can take only two values $\pm 1$. $J$ is the interaction energy between nearest-neighbor spins $\langle i,j\rangle$, and $H$ is the external magnetic field. The magnetization $M$ (the order parameter) is
\begin{flalign}\label{order parameter}
&\qquad M=\frac{1}{V}{\langle \sum_{i}{s_{i}}\rangle}=\frac{s}{V},&
\end{flalign}
where $s=\sum_{i}{s_{i}}$ and $V=L^d$ denote the total spin and volume of the lattice, respectively, where $d=3$ is the dimension of the lattice. The magnetization is a function of the external magnetic field $H$ and the reduced temperature $t=(T-T_c)/{T_c}$, where $T_c$ is the critical temperature. The reduced temperature describes the distance away from the critical point. At $t>0$, it is the crossover side. At $t<0$, it is the first order phase transition side.

The high-order cumulants of the order parameter can be obtained from the derivatives of magnetization with respect to the external magnetic field $H$ at fixed $t$,
\begin{flalign}\label{cumulants}
&\qquad \left.\kappa_{n}(t,H)=(\frac{\partial^{n-1} M}{\partial H^{n-1}})\right|_{t}.&
\end{flalign}
In particular, the sixth order cumulant is as follows,
\begin{flalign}\label{sixth cumulant}
& ~\left.\kappa_{6}(t,H)=(\frac{\partial^{5} M}{\partial H^{5}})\right|_{t} & \nonumber \\
&~=\frac{1}{V}(\langle\delta s^6\rangle-10\langle\delta s^3\rangle^2+30\langle\delta s^2\rangle^3-15\langle\delta s^4\rangle\langle\delta s^2\rangle),&
\end{flalign}
where $\delta s=s-\langle s\rangle$.

To study the behavior of the high-order cumulants, one way is to simulate the Ising model by the Monte Carlo method. Another way is to use a parametric representation of the Ising model. In this representation, the magnetization $M$ and reduced temperature $t$ can be parameterized by two variables $R$ and $\theta$~\cite{linear para, linear para 3},
\begin{flalign}\label{parametric}
&\qquad M=R^{\beta}\theta,~~~~~~t=R(1-\theta^2).&
\end{flalign}
The equation of state of the three-dimensional Ising model can be given by the parametric representation in terms of $R$ and $\theta$ as
\begin{flalign}\label{equation state}
&\qquad H=R^{\beta\delta}h(\theta),&
\end{flalign}
where $\beta$ and $\delta$ are critical exponents of the three-dimensional Ising universality class with values 0.3267(10) and 4.786(14), respectively~\cite{Ising exponents}.

If $M$, $t$ and $h$ are analytic functions of $\theta$, the analytic properties of the equation of state are satisfied~\cite{linear para 1}. The analytic expression of the high-order cumulants can be derived in the parametric representation.

In fact, the function $h(\theta)$ is analytic and an odd function of $\theta$ because the magnetization is an odd function of the magnetic field $M(-H)=-M(H)$. $h(\theta)=0$ at $\theta=0$ corresponds to the crossover line $H=0$, $T>T_c$. It vanishes for another $\theta=\theta_0$ corresponds to the coexistence curve $H=0$, $T<T_c$ (the first order phase transition line).

One simple function of $h(\theta)$ obeying all the demands is as follows,
\begin{flalign}\label{equation h}
&\qquad h(\theta)=\theta(3-2\theta^{2}).&
\end{flalign}
This is an exact representation of the equation of state of the Ising model to order $\varepsilon^{2}$, where $\varepsilon$ is a parameter related to the number of dimensions of space. $\varepsilon$-expansion is one of the techniques to explore the critical phenomena. It is enough for our purpose, although the parametric representation can be exact up to order $\varepsilon^{3}$~\cite{linear para 3}. On the other hand, there is an excellent agreement of the scaling magnetization data from Monte Carlo simulations and the equation of state in parametric representations~\cite{J. Engels}.

Using Eq.~\eqref{sixth cumulant} to Eq.~\eqref{equation h}, the sixth order cumulant can be expressed in the parametric representation. For our purposes, it would be enough to take the approximate values $1/3$ and $5$ for the critical exponents $\beta$ and $\delta$, respectively. Then the sixth order cumulant can be expressed as follows,
\begin{flalign}\label{six in para}
&\qquad\kappa_{6}(t,H)
=240\frac{\sum_{n=0}^{n=9}{a_{2n}\theta^{2n}}}{R^8(\theta^2-3)^7(2\theta^2+3)^9},&
\end{flalign}
where $a_{2n}$ are the expansion coefficients. Their values are listed in Table I.

\begin{center}
\tabcaption{ \label{tab1}  Values for the expansion coefficient $a_{2n}$.}
\footnotesize
\begin{tabular*}{80mm}{c@{\extracolsep{\fill}}ccccc}
\hline
$a_0$ ~~~~~~~&$a_2$ &$a_4$ &$a_6$ & $a_8$ \\
\hline
-98415\hphantom{00}~~ & 3306744~ & -11234619~ & 7120872~&-2736261~\\
\hline
$a_{10}$~~~~~~~ & $a_{12}$ & $a_{14}$ & $a_{16}$ & $a_{18}$\\
\hline
501120\hphantom{00}~~&-53001~&1560~&-8~&8 \\
\bottomrule
\end{tabular*}
\end{center}

The main difference in the two ways is that the results of the high-order cumulants are from finite systems by the Monte Carlo simulation, while in the parametric representation, the results are under the condition of the thermodynamic limit.

\section{Behavior of the sixth order cumulant in the parametric representation}

In Eq.~\eqref{six in para}, the sixth order cumulant has two zero values at $\theta=0.5723$ and $\theta=0.183$, respectively. Combining Eqs.~\eqref{parametric}, \eqref{equation state} and Eq.~\eqref{equation h}, the reduced temperature $t$ can be expressed as a function of $\theta$ and the external magnetic field $H$,
\begin{flalign}\label{th}
&\qquad t=\frac{1-\theta^2}{(\theta(3-2\theta^{2}))^{1/(\beta\delta)}}H^{1/(\beta\delta)}.&
\end{flalign}
Then we can get the sign distribution of $\kappa_6$ on the $t-H$ plane. We show it as a density plot in Fig.~1(a).

\begin{center}
\includegraphics[width=7.5cm]{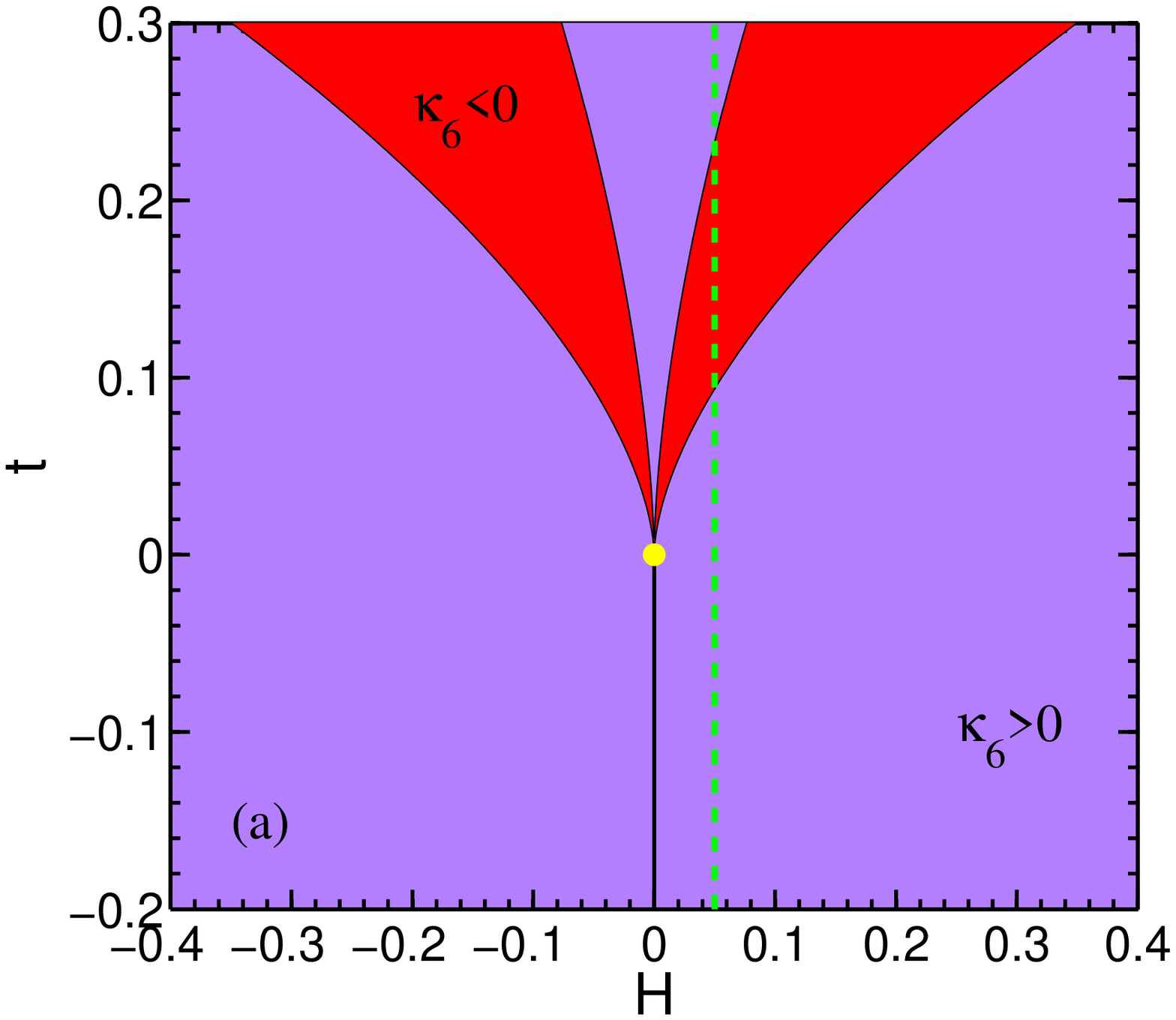}
\includegraphics[width=7.5cm]{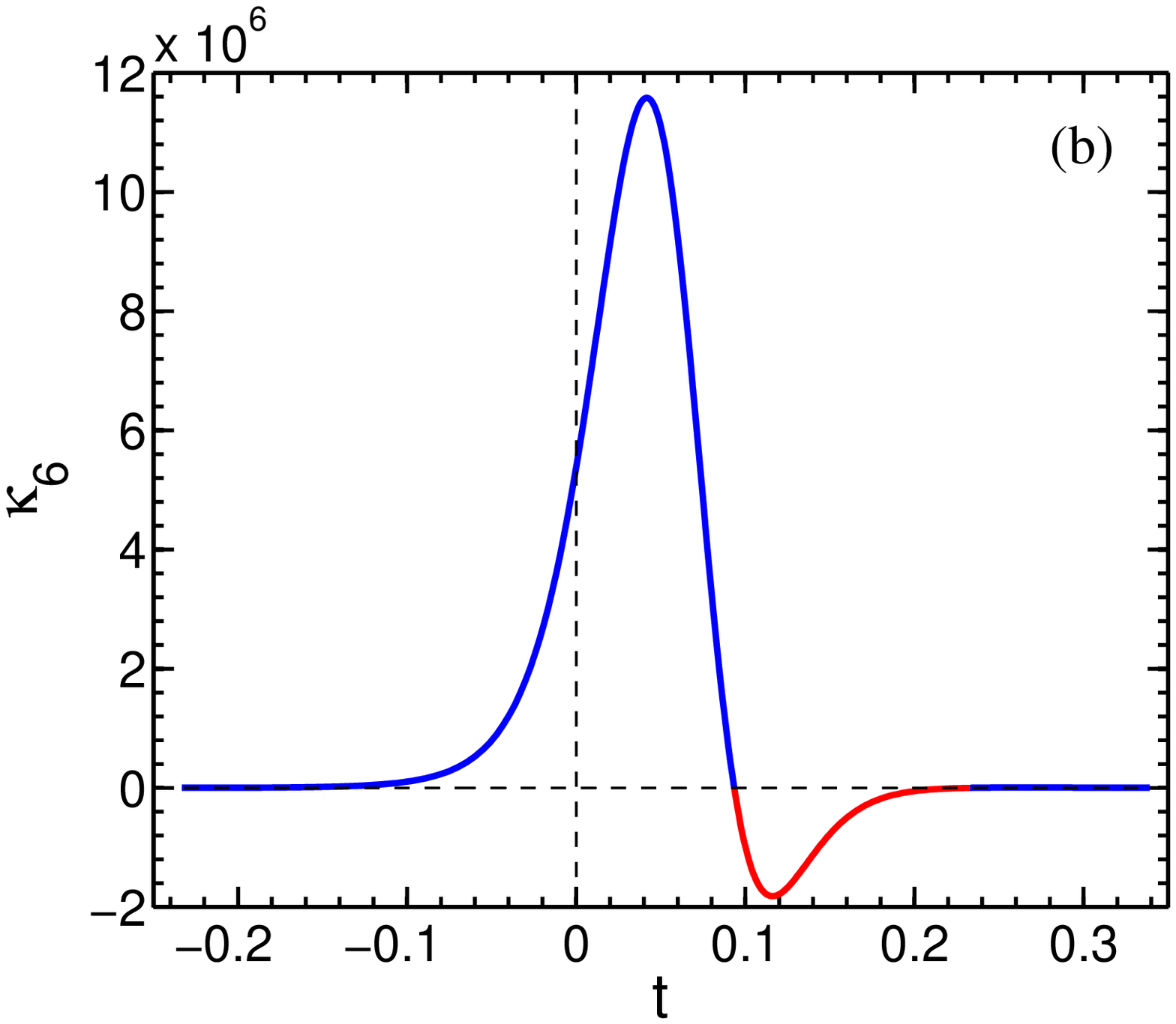}
\vspace{4pt}
\figcaption{\label{fig1}(color online) (a) Sign distribution of the values of $\kappa_{6}$ given by Eq.~\eqref{th} on the $t-H$ plane. (b) The temperature dependence of $\kappa_{6}$ along the green dashed line in sub-figure (a).}
\end{center}

In Fig.~1(a), the yellow point at $H=0$ and $t=0$ indicates the critical point of the three-dimensional Ising model. On the black curves, the values of the sixth order cumulant are zeros. The middle two black curves correspond to $\theta=0.183$. The other two correspond to $\theta=0.5723$. In the red regions, the sixth order cumulant takes negative values. It is positive in the purple region. Specially, the red region is separated into two parts by the purple region, which is different from the sign distribution of the fourth order cumulant given in Ref.~\cite{Stephanov-prl107}. Although the values of the sixth order cumulant are positive in the smaller purple region between the two red regions, it approximately equals to zero.

Along the green dashed line in Fig.~1(a), we study the temperature dependence of the sixth order cumulant at fixed external magnetic field, as shown in Fig.~1(b). The blue line shows where the sixth order cumulant is positive and the red line shows where it is negative. It is clear that the sixth order cumulant is positive but approximately zero at the higher temperature side, which corresponds to the green dashed line above the red region in Fig.~1(a). Then as the temperature decreases and approaches the critical point, the sixth order cumulant has a negative valley.

At fixed external magnetic field, the qualitative temperature dependence of the sixth order cumulant is similar to the parametric representation and Monte Carlo simulations~\cite{NPA}. Quantitatively, in different sizes of systems, due to the finite-size effects, the temperature of the sign change of the sixth order cumulant will be different.

\section{Finite-size effects on the sixth order cumulant}

Using Monte Carlo simulations, we study the finite-size effects on the temperature of the sign change of the sixth order cumulant. The simulations are performed on three different sizes of lattices with $L=12, 14, 16$ at 11 different external magnetic fields between $H=0.005$ and $H=0.04$ by the Wolff cluster algorithm~\cite{wolff}. The helical boundary conditions are used. Because of the symmetry of the spin inversion, with negative external magnetic fields the values of the temperatures of the sign change are just the same as those of the corresponding positive ones.

On the $t-H$ plane, we show the positions of the first sign changes at lower temperatures of the sixth order cumulant (e.g., the first zero at lower temperature in Fig.~1(b)) by the points, as shown in Fig.~2. The positions of the first sign changes are closer to the critical point and easier to be distinguished in the Monte Carlo simulations. The black points, blue squares and red stars present the results on lattices of size $L=12, 14, 16$, respectively. Specially, pink triangles present the results from the parametric expression of the sixth order cumulant in the parametric representation.

In Fig.~2, for each size, below the points, the values of the sixth order cumulant are positive; above the points, the values are negative. It is clear that, at a fixed external magnetic field, the temperature of sign change increases as the size increases. The amplitude of increase decreases with increasing $H$. The bigger the external magnetic field, the smaller of the size of the system to get saturation. That is to say, the features of the thermodynamical quantities do not depend on the system size when it is bigger than some value. In fact, when it deviates from the critical point, the correlation length decreases. The system size needed to get saturation decreases, too.

\begin{center}
\includegraphics[width=7cm]{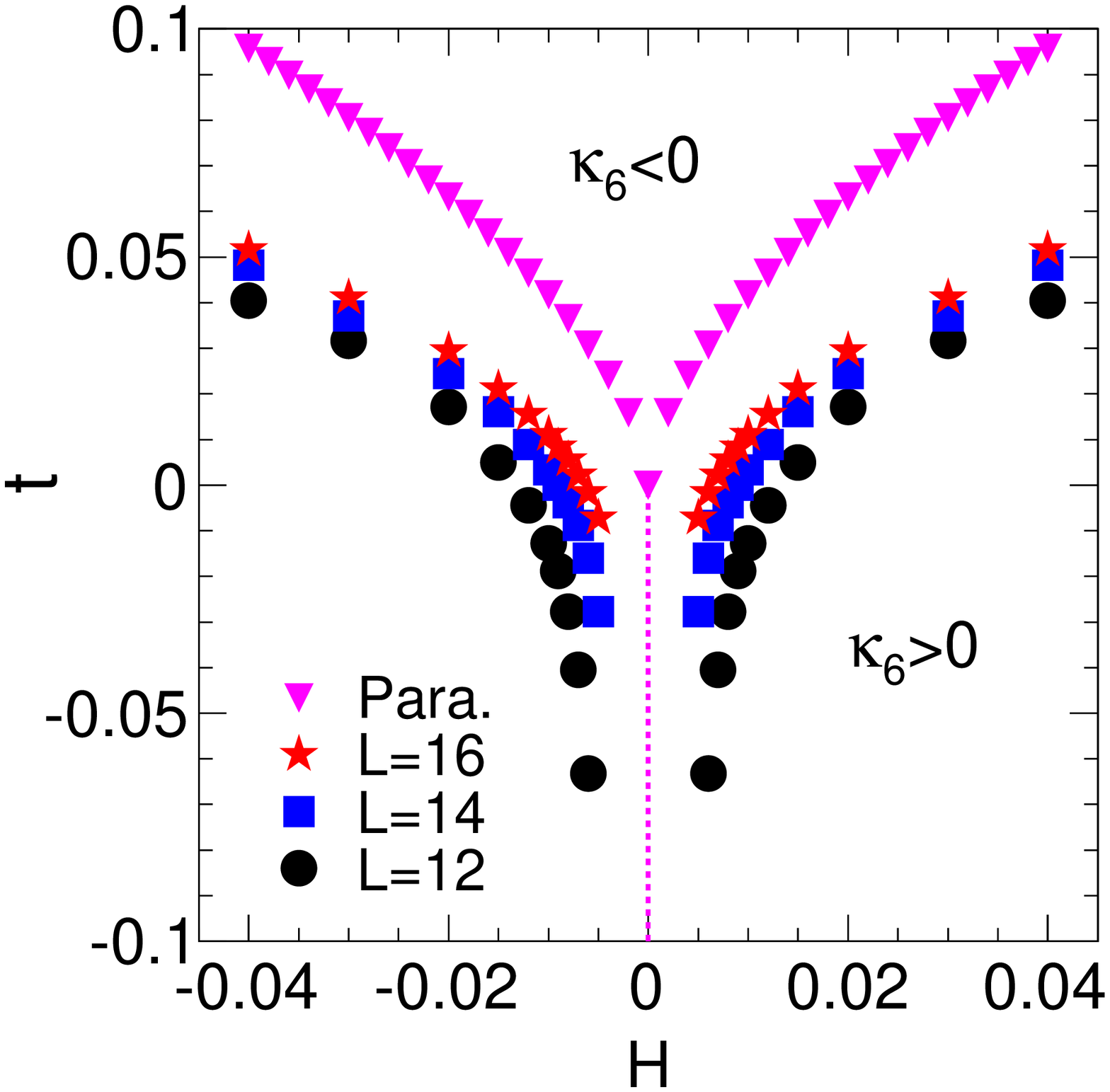}
\vspace{4pt}
\figcaption{\label{fig2}(color online) The finite-size effects on the temperature of the first sign change of $\kappa_{6}$ on the $t-H$ plane in the three-dimensional Ising model. Para. means pink triangles are results from the parametrization representation.}
\end{center}

According to the results from the parametric expression of the sixth order cumulant, i.e. the pink triangles in Fig.~2, as the external field decreases to zero, the temperature of the sign change reduces to the critical temperature. At the critical point, the sixth order cumulant is divergent. In a finite system of size $L=12$, as the external field decreases, the temperature of the sign change decreases. According to the trend, it can be inferred that the temperature of the sign change is smaller than the critical temperature at vanishing external field. As the system size increases from $L=12$ to $L=16$, the pseudo-critical temperature (here we mean the temperature of the sign change at $H=0$ in a finite system) increases. When reaching the thermodynamical limit, the pseudo-critical temperature will approach the real critical temperature. The qualitative result is consistent with the linear sigma model with constituent quarks and the two flavor quark-meson model~\cite{finitesize1, finitesize2, finitesize3, finitesize4}, which inferred that the QCD pseudo-critical point should shift to the higher temperature side as the size of the system increases.

Comparing the behavior of the sixth order cumulant with the fourth order one, e.g., Fig.~1(b) in this paper and Fig.~1(b) in Ref.~\cite{Stephanov-prl107}, qualitatively, at a fixed $H$, their trends varying with temperature are similar. When the temperature is approached from the higher temperature side, the sixth and fourth order cumulants are both negative. Quantitatively, the negative valley is more obvious in the sixth order cumulant than the fourth order one. For a constant $H$, the ratio of the maximum to the minimum of the sixth order cumulant is a constant. According to Eq.~\eqref{six in para}, it approximates -6. For the fourth order cumulant, the ratio is approximately -28. From this result, the benefit of using the sixth order cumulant to search for the critical point is obvious.

On the other hand, a high price is needed to calculate the sixth order cumulant. First, the higher the order of the cumulant, the more statistics are needed. Second,  in the vicinity of the critical point, the $n-th$ order cumulant depends on the system size by $L^{n(d-\beta/\nu)}$~\cite{Privman}. The higher the order of the cumulant, the more serious the finite-size effects.

\section{Summary}

Taking advantage of the parametric representation of the three-dimensional Ising model, we have studied the sign distribution of the sixth order cumulant of the order parameter in the $t-H$ plane. At a fixed external magnetic field, we calculated the temperature dependence of the sixth order cumulant. We found that at nonzero external magnetic field, when the critical point is approached from the higher temperature side, the sixth order cumulant has a negative valley. The width of the negative valley narrows
with decreasing external field.

Through Monte Carlo simulations, the finite-size effects on the temperature of the sign change of the sixth order cumulant have been studied. At the same external magnetic field, the temperature of the sign change of the sixth order cumulant increases with the increasing size of the system in the Ising universality class. This indicates that the pseudo-critical point should shift towards higher temperatures with increasing size. The trend is consistent with that of the linear sigma model with constituent quarks and the two flavor quark-meson model.

Results of the sixth order cumulant are compared with the fourth order one. At fixed external magnetic field, qualitatively, their temperature dependence is similar. Quantitatively, the negative valley in the sixth order cumulant is more obvious. At the same time, the finite-size effects are also more serious in the higher order of cumulant.

\end{multicols}

\clearpage

\end{CJK*}
\end{document}